# Dual MINE-based Neural Secure Communications under Gaussian Wiretap Channel

Jingjing Li*, Zhuo Sun*, Lei Zhang†, Hongyu Zhu‡
* Key Laboratory of Universal Wireless Communications, Beijing University of Posts and Telecommunications Communications
†State Grid Fujian Electric Power Co.LTD, Fuzhou, China
‡State Grid Hunan Electric Power Co.LTD, Changsha, China
Email:{jingjing, zhuosun}@bupt.edu.cn, {jingjing, zhuosun}, 190398281@qq.com, zhuhongyu51396@163.com

*Abstract*—Recently, some researches are devoted to the topic of end-to-end learning a physical layer secure communication system based on autoencoder under Gaussian wiretap channel. However, in those works, the reliability and security of the encoder model were learned through necessary decoding outputs of not only legitimate receiver but also the eavesdropper. In fact, the assumption of known eavesdropper's decoder or its output is not practical. To address this issue, in this paper we propose a dual mutual information neural estimation (MINE) based neural secure communications model. The security constraints of this method is constructed only with the input and output signal samples of the legal and eavesdropper channels and benefit that training the encoder is completely independent of the decoder. Moreover, since the design of secure coding does not rely on the eavesdropper's decoding results, the security performance would not be affected by the eavesdropper's decoding means. Numerical results show that the performance of our model is guaranteed whether the eavesdropper learns the decoder himself or uses the legal decoder.

## I. Introduction

With the development of 5G technology, the application of wireless communication technology goes deep into all fields of society. A variety of confidential and sensitive data in wireless networks are growing rapidly, which leads to prominent information security issues. Compared with traditional wired networks, wireless communication is vulnerable to eavesdropping, attacks, and interference due to the openness of the transmission medium and the characteristics of broadcasting. Most of the traditional security mechanisms are based on cryptography. On the one hand, with the increasing performance of processors, the rapid improvement of computer computing capacity, and the maturity of distributed computing theory, the classical encryption algorithm system based on complexity is becoming increasingly unstable. On the other hand, the appearance of quantum computer makes it more vulnerable and easier to be cracked. Therefore, physical layer security technology is triggering wide interest in the academic field, which process transmitted signals or implement corresponding security policies using channel information with the characteristics of lightweight, high security, and can communicate safely without using the pre-shared key.

A considerable research on physical layer security is based on information theory. Researchers make a profound study on various wireless wiretap channel models, and seek ways to make the security transmission rate as close as possible to security capacity. For example, some typical methods include directional modulation [1]-[3], polarization modulation [4] and cooperative interference [5]-[7]. Through beamforming, artificial noise, depolarization effect or the introduction of cooperative jammer to send artificial interference, the eavesdropper cannot demodulate useful information due to the serious deterioration of the received signal, thus improving the security of the system in the physical layer. In this paper, we also start from the perspective of information theory, maximizing the secure transmission rate of the system based on the wiretap channel model.

Deep neural networks (DNNs) have been successfully applied in many fields, such as computer vision [1], natural language processing [2], and speech recognition [3]. When it comes to using DNNs to learn the physical layer security communication method, the autoencoder is the first attempt. In [4], the author first proposed to apply deep learning to the communication system, and an autoencoder structure was used to construct a communication system model. The model greatly improved the reliability of communication through end-to-end training. An application example of autoencoder in the wiretap channel can be found in [5] which considers the encoding and decoding scheme of the Gaussian wiretap channel with a single antenna. The main idea is to construct an end-to-end communication system model, and optimize the encoder and decoder jointly to achieve a balance between reliability and security. In our previous work [6], we also used end-to-end learning to solve two communication security problems: confidential transmission and user authentication. However, learning the end-to-end communication system is a challenging task. It is necessary to establish a differentiable channel model to meet the reverse transfer of parameters in the process of training the neural networks, which is impossible in real-world channel conditions.

One way to dissolve the above issue is to utilize mutual information estimation [7]. Using this method, it was shown in [8], that a secure encoding can be learned based on a mixed loss function that consists of the approximate mutual information of the legitimate channel, learned from signal samples, and the cross-entropy function from the clustering enforcing the fake distribution. As a result, the intermediate steps of estimating and generating channel distribution can be skipped. However, the decoding results of the eavesdropper in their mixing function are difficult to obtain in the actual environment, especially in passive eavesdropping scenario. Besides, they only use one eavesdropping structure to verify the security of the system, which led to an ideal result.

In this paper, we present a dual MINE-based neural secure communications model, two MINE modules are used to estimate the channel capacity of the legitimate channel and eavesdropper channel respectively, the estimated mutual information are subsequently used to construct the security capacity constraint as the objective function for training the encoder. Our main contributions and novelties are summarized as follows: (1) the encoder and decoder are completely independently training, avoiding the problem of differentiable channel model needed. (2) We can train the encoder reliably just with the signal samples of the main channel and eavesdropper channel, enabling deep learning for encoder without explicit knowledge of decoders. (3) We provide eavesdropper the two most possible decoder options to verify the system performance. One is that the eavesdropper knows the encoder used by the transmitter, and learns a corresponding decoder model by himself; Another is that the eavesdropper can obtain the decoder of the legitimate receiver and uses it to decode its received signals.

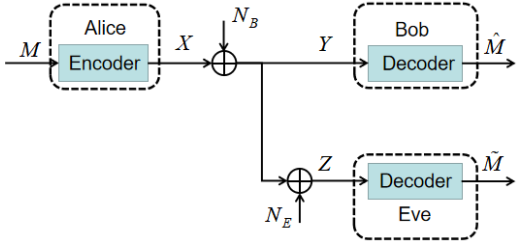

Fig. 1: Degraded Gaussian wiretap channel. The sender (Alice) sends confidential information $M$ to the legitimate receiver (Bob) while keeping the external eavesdropper (Eve) unaware of it.

The rest of this paper is organized as follows. We present the basis of the Gaussian wiretap channel used in our work in Section II. The implementation details of the proposed dual MINE-based neural secure communications model and the training methods of each block are detailed in Section III. We provide the numerical results and correlation analysis in Section IV. Finally, the paper is concluded in Section V.

*Notations*—Random variables of all dimensions are denoted in boldface upper case letters, e.g., $M$. The sets of codebooks are denoted with calligraphic letters, i.e. $\mathcal{M}$. The dimension n is marked in the upper right corner of the letter. e.g., $X^n$. The element $i$ is marked in the lower right corner of the letter (e.g., $X_i$). The probability mass or density function of event $x$ is denoted as $p(x)$. $\mathbb{R}(\mathbb{C})$ is the set of real numbers (complex numbers). The expectation is denoted by $E\{\cdot\}$.

## II. BASIS OF GAUSSIAN WIRETAP CHANNEL

The simplest scenario which involves both tasks of reliable transmission and secrecy is the wiretap channel [9], which is a three-node network consists of a transmitter, a legitimate receiver, and an eavesdropper. The goal of the transmitter is to communicate reliably with the legitimate receiver while keeping the eavesdropper as ignorant of the confidential information as possible. In this paper, we consider the degraded Gaussian wiretap channel extended from Wyner's wiretap channel [10] as depicted in Fig. 1. It is a degraded broadcast channel with memoryless additive Gaussian white noise [11]. It is believed that this will lay a foundation for us to understand and study more complex communication scenarios.

The procedures to transmit confidential information mainly consist of three phases, which are summarized as follows:

First, Alice transmits a message $M$ from the set of $\mathcal{M} = \{1,2,\dots,2^{nR}\}$ at a rate $R$ to the legitimate user Bob. Alice encodes it into a codeword of block length $n$ using an encoding function $f(M) = x^n(M) \in \mathbb{C}^n$. Besides, we assume that the variance of the transmitted codewords is not allowed to exceed a certain value $P$

$$E\{x(M)^2\} \leq P, \ \forall \in \mathcal{M} \quad (1)$$

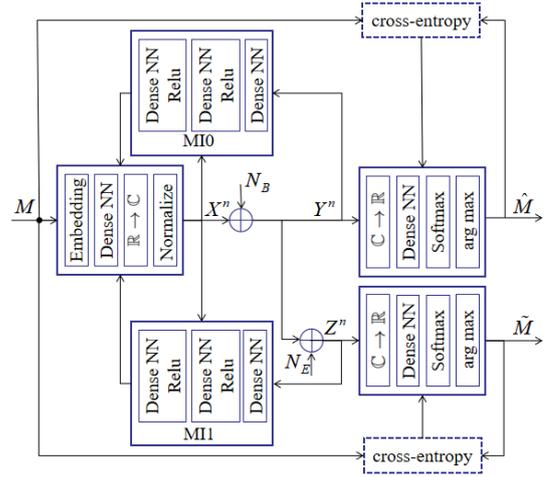

Fig. 2: The architecture of Dual MINE-based Neural Secure Communications model.

Second, the main channel between Alice and Bob and the eavesdropper channel between Alice and Eve are defined by

$$Y_i = X_i + N_{B,i} \ \ for \ i \in \{1,\dots,n\}, N_{B,i} \sim \mathcal{CN}(0,\sigma_B^2) \quad (2)$$

$$Z_i = Y_i + N_{E,i} \ \ for \ i \in \{1,\dots,n\}, N_{E,i} \sim \mathcal{CN}(0,\sigma_E^2) \quad (3)$$

where $X_i, Y_i, Z_i$ denote the channel input, Bob's observation, and Eve's observation, respectively. In the following, we consider the degraded case such that $\sigma_B^2 < \sigma_E^2$.

The third part is based on the premise that the eavesdropper knows the encoder of the transmitter, and then builds a decoder neural network and trains it by himself. The task of Bob and Eve is to estimate and recover the original message using the decoding functions

$$g(y^n) = \widehat{M} \quad (4)$$

$$h(z^n) = \widetilde{M} \quad (5)$$

The symbol error rate (SER) of the two receivers is then defined as $P_B = E\{\widehat{M} \neq M | Y\}$, $P_E = E\{\widetilde{M} \neq M | Z\}$, respectively.

## III. DUAL MINE-BASED NEURAL SECURE COMMUNICATIONS MODEL

Aiming at provide a secure communication over the degraded Gaussian wiretap channel, we build a fully learning based communication system shown in Fig. 2, of which all the modules are implemented with DNNs, including an encoder block, two mutual information estimators, and two decoder blocks. Note that the encoder block and the decoder block are independent and not constructed as autoencoder model. As a result, the intermediate process of passing the gradient back from the decoder through the channel can be skipped, thus avoiding the need of differentiable channel model.

Wiretap channel security capacity is the basis of physical layer security technology research, which plays an important role in guiding the physical layer security mechanism. Therefore, we consider to use the security capacity as a metric to learn the optimal encoding function of the Gaussian wiretap channel.

The secrecy capacity of a Gaussian wiretap channel was derived in [12] and was found to be the capacity difference between the main and the eavesdropper channels, i.e.,

$$C_s = \max_{p(x)}\{C_m - C_w\} \quad (6)$$

where $C_m$ is the main channel capacity and $C_w$ is the eavesdropper channel capacity. And the capacity $C$ for a Gaussian channel is known to be $C = \max_{p(x)} I(X;Y)$. Therefore, the secrecy capacity calculation further simplifies to just the difference in the mutual information of two channels.

$$C_s = \max_{p(x)}[I(X;Y) - I(X;Z)] \quad (7)$$

The two MINE modules of our proposed model are used to achieve the estimation of the mutual information. Based on this, the encoder loss function can satisfy the reliability and security constraints.

*A. MINE Block*

As described above, we use the mutual information between input-output measurements of the wireless channel to optimize the encoder weight. In this work, we choose to utilize MINE to achieve this goal, which using SGD and Donsker-Varadhan representation of the Kullback-Leibler (KL) divergence defined as

$$I(A;B) = D_{KL}(p(a,b)\|p(a)p(b)) \geq \sup_{\theta \in \Theta} E_{p(a,b)}\{T_\theta(A,B)\} - \log E_{p(a)p(b)}\{e^{T_\theta(A,B)}\} \quad (8)$$

where $A$ and $B$ denote two random variables with arbitrary distributions $p(a), p(b)$, respectively. In the MINE, MI0 and MI1 represent two equally constructed neural networks, each of which has a fully-connected hidden layer with a relu activation function, and a linear output layer.

The specific operation of calculating mutual information is explained by taking MI0 as an example. We first sample the joint distribution $p(x^n, y^n)$ and the marginal distributions $p(x^n)$ and $p(y^n)$, and then send the results to MI0 to get the $T_\theta$. Finally, the mean value of $T_\theta$ is calculated to approximate the expectations. Given that $T_\theta$ is expressive enough, the above lower bound converges to the true mutual information. Hence the mutual information for $X^n$ and $Y^n$ is calculated as

$$\tilde{I}_\theta(X^n; Y^n) := \frac{1}{k}\sum_{i=1}^{k}[T_\theta(x_{(i)}^n, y_{(i)}^n)] - \log \frac{1}{k}\sum_{i=1}^{k}\left[e^{T_\theta(x_{(i)}^n, \bar{y}_{(i)}^n)}\right] \quad (9)$$

where the $k = 64$ represents the number of samples. The objective functions for optimization MI0 and MI1 are then set as $\max_\theta \tilde{I}_\theta(X^n; Y^n)$ and $\max_\vartheta \tilde{I}_\vartheta(X^n; Z^n)$, respectively.

*B. Encoder Block*

The encoder consists of an embedding followed by a dense hidden layer with an elu activation function, and a linear output layer. Its 2n-dimensional real-valued output is converted to an n-dimensional complex-valued vector and the power constraint is applied to the result to form $X^n$.

The purpose of training encoder is to minimize SER and information leakage. This is a multi-objective programming problem (MOP) that can be solved by combining multiple objectives into a scalar objective. Here we use the weighted sum and scalar objective for the objective function, expressed as

$$\max_\varphi \alpha I(X_\varphi^n(M); Y^n) - (1-\alpha)I(X_\varphi^n(M); Z^n) \quad (10)$$

where the value of $I(X_\varphi^n(M); Y^n)$ will affect the reliability of the system, while the value of $I(X_\varphi^n(M); Z^n)$ will affect the security of the system. The α in (10) is a specific weight that can be used to balance two objectives.

*C. Decoder Block*

The structure of the decoder is almost symmetrical to the encoder, except that a softmax activation function is used in the final dense layer. Its output is a probability vector that assigns a probability to each of the possible messages. Then we use an arg max layer to select the index of the maximum value in the probability vector, which represents the information value $\hat{M}$ of the decoding result. The target of training Bob and Eve is to decode secure coded signals, that is, to restore the transmitted message as much as possible. In this case, the cross-entropy loss function with variants of stochastic gradient descent [13] is a satisfactory choice. The loss function of the two decoders can be written as

$$L_B := H(M, \hat{M}) = -\frac{1}{k}\sum_{i=1}^{k} \log p_{\hat{M}} \quad (11)$$

$$L_E := H(M, \tilde{M}) = -\frac{1}{k}\sum_{i=1}^{k} \log p_{\tilde{M}} \quad (12)$$

where $k = 500$ represents the number of samples, $p_{\hat{M}}$ and $p_{\tilde{M}}$ denote the estimated probability of the message received by the two receivers, respectively. They describe the distance between the actual output and the expected output, that is, the smaller the cross-entropy, the smaller the SER.

Our training procedure is divided into three phases. Firstly, we train the MI estimation network $T_\theta$ for 5 epochs with 500 iterations. At this point, the parameters of the encoder network are only randomly initialized and the estimated mutual information might not reflect the final estimated value. Secondly, the security of the system is learned by training the encoder and two estimators iteratively. The parameters of one model will be fixed while training the other so that the optimal point can be found. Finally, we train the legal decoder to achieve reliable communication with the transmitter and train the eavesdropper's decoder to verify the security of the system.

Note that, all neural networks in the system are optimized by Nadam optimizer [14] with a learning rate of 0.001. The signal-to-noise ratio (SNR) of both channels is fixed to 7dB. And the output of the encoder network has a unit average power normalization.

IV. NUMERICAL RESULTS AND ANALYSIS

*A. Reliability and Security*

In this part, we test the system under different $SNRs \in (0dB, 21dB)$ of the legal channel. Firstly, we set the SNR of the eavesdropper channel to 7dB. The SER curves can be seen in Fig. 3. The 16QAM is taken as the benchmark for comparison. One can see that the reliability performance of our method is close to that of 16QAM at low SNR, and the reliability performance gap is within 1dB when SNR gets higher. This is due to the trade-off between reliability and security we considered, the system sacrifices part of reliability for security. Moreover, the SER of Eve is limited to a large value even at high SNR. Subsequently, we conducted a test to see if Eve could achieve the purpose of eavesdropping using Bob's decoding network. The result is shown in Fig. 4. As expected, the decoding result of Eve has a high symbol error rate, which proves that the model achieves presentable security regardless of the eavesdropping means.

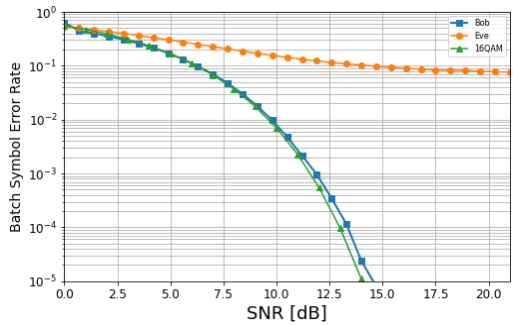

Fig. 3: SER performance of the proposed method for a 16-dimensional codeword constellation. The weight $\alpha$ in (10) is 0.7. Eve uses its own trained neural network as the decoder. A theoretical estimation of 16QAM as a reference.

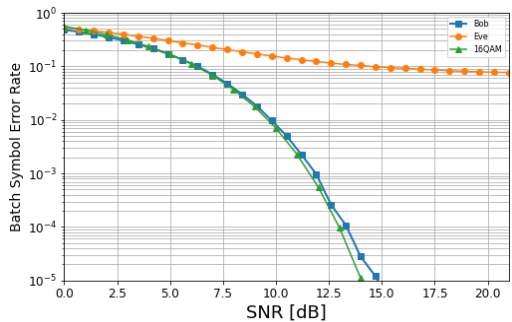

Fig. 4: SER performance of the proposed method for a 16-dimensional codeword constellation. The weight $\alpha$ in (10) is 0.7. Eve uses Bob's neural network as the decoder. A theoretical estimation of 16QAM as a reference.

From Fig. 5, we can find that the coefficient in (10) is the most important factor to determine the system performance. The SER of Bob and Eve get smaller with increasing weight α, which suggests that the NN can find better constellations. The drawback is that the security is reduced, due to the trade-off between security and reliability. Therefore, we have taken a conservative approach, which resulted in $\alpha = 0.7$.

### B. Comparison to Existing Methods

To compare our approach with others presented in [5] and [8], we reproduced a system model similar to the one they used. The most prominent feature of this model is the use of coset coding algorithm. The actual message is hidden in the cosets. Bob can distinguish the real codeword from the cosets, while Eve can only recognize the cosets of the codeword. The general process is that the input signal $M$ is coded into $S$ by one-hot encoding, and then the $S$ is encoded by coset to get equalized input symbol distribution $\bar{S}$. The decoding distribution of Bob and Eve is expressed as $\hat{S}$ and $\tilde{S}$, respectively.

For comparison, we use the following three methods to implement the security of communication system:

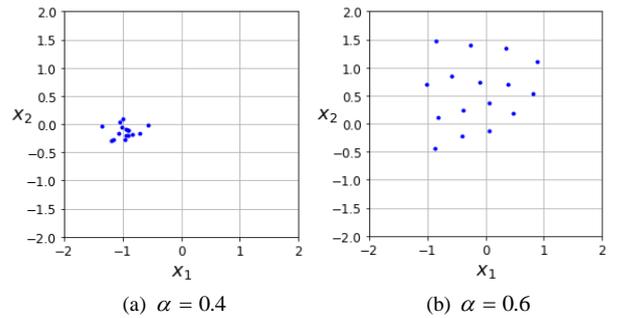

(a) $\alpha = 0.4$      (b) $\alpha = 0.6$

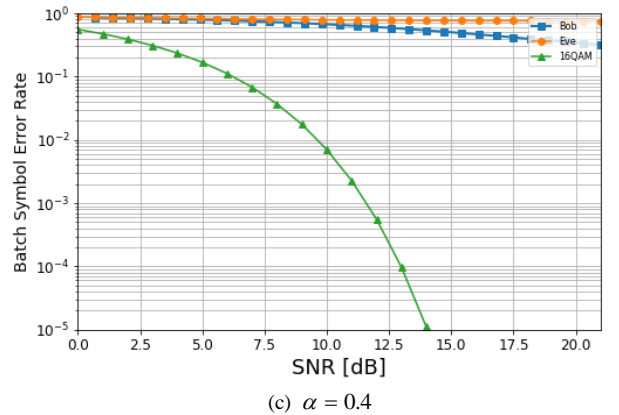

(c) $\alpha = 0.4$

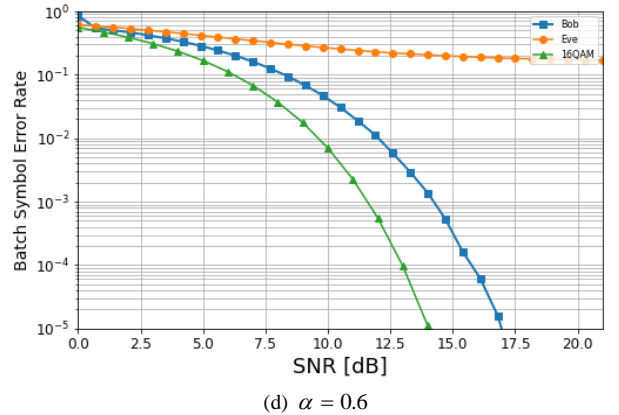

(d) $\alpha = 0.6$

Fig. 5: The resulting encoding constellations are shown for 16 symbols based on: (a) the encoder training weight $\alpha = 0.4$ ; (b) the encoder training weight $\alpha = 0.6$ . Their corresponding SER performance is shown in (c) and (d), respectively.

*1) Autoencoder and Cross-Entropy (AE+CE):* This is the method used in [5]. They use an end-to-end approach to jointly train the encoder and decoder with a secure loss function based on cross-entropy which is defined as

$$L_{sec} := \alpha H(M;\hat{M}) + (1-\alpha) H(\bar{M};\tilde{M}) = \\ -\alpha \sum_{i=1}^{|\mathcal{M}|} S_i \log \hat{S}_i - (1-\alpha) \sum_{i=1}^{|\mathcal{M}|} S_i \log \tilde{S}_i \quad (13)$$

*2) Mutual Information and Cross-Entropy (MI+CE):* This is the method raised in [8]. The training of encoder and decoder is carried out separately, and the security loss function used in training encoder is the combination of mutual information and cross-entropy, i.i.d.

$$L_{sec} := \alpha I(X^n(M); Y^n) - (1-\alpha) H(M;\tilde{M}) \quad (14)$$

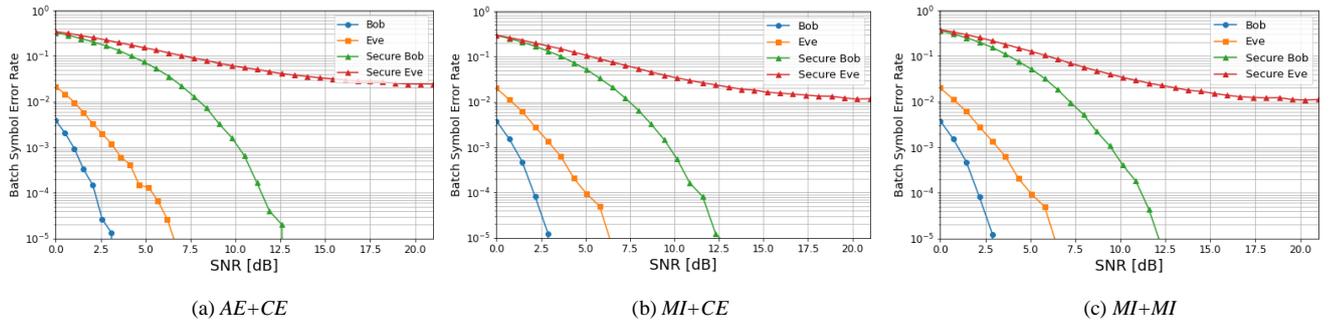

(a) *AE+CE*     (b) *MI+CE*     (c) *MI+MI*

Fig. 6: Shows the SER for the three methods. Bob and Eve indicate the error rate of the codeword transmitted before the security encoding, and correspondingly, Secure Bob and Secure Eve indicate the error rate after the security encoding, reflecting the reliability and security of the system respectively.

where $H(M, \widetilde{M})$ is the same as the one in (13).

*3) Mutual Information and Mutual Information (MI+MI):* This is the Dual MINE-based Neural Secure Communications model proposed in this paper which also trains the encoder and decoder separately. Moreover, compared with method 2, our security constraints can be implemented without knowing the output of the decoder. Therefore, the safety loss is as follows:

$$L_{sec} := \alpha I(X^n(M); Y^n) - (1-\alpha) I(X^n(M); Z^n) \quad (15)$$

Fig. 6 shows the performance of the three methods in security encoding. We can analyze the reliability and security of each method through the green and red curves. One can see that, Fig. 6c has a better reliability in the case of similar security as Fig. 6b. This shows that our method can achieve good performance even without prior knowledge of the decoder. By comparing Fig. 6a and Fig. 6b, it can be found that the former has better security in the case of similar reliability. This is because the latter two methods train the encoder and decoder separately which cannot ascertain the global optimality of the system, even if they all achieve their respective optimal performance. But as described in the first section, this training method avoids the dependence on a differentiable channel model.

## V. CONCLUSIONS

In this paper, we contribute to give a model based on two mutual information estimators for secure communication in the degraded Gaussian wiretap channel where each party is equipped with DNNs. The encoder is trained by security capacity in a manner to provide reliability for legal decoder while ensuring security against wiretapper. The presented results are compared with the other two security constraint methods based on deep learning which highlights the superiority of our proposed method in robustness to different eavesdropping environments without prior knowledge of the decoder. However, future research needs to investigate the performance under unknown eavesdropping channel conditions. Moreover, we can continue to explore the direction corresponds to the case where the broadcast channel is no longer degraded.

## ACKNOWLEDGEMENT

This paper is supported by "2020 Industrial Internet Innovation and Development Project (Project No. TC200H01L)".